\documentclass[twocolumn]{aastex63}
\usepackage{amssymb,amsmath,amstext,natbib}
\accepted{by AJ on Oct, 2021}

\graphicspath{{./}{figures/}}

\shorttitle{{\it Spitzer}/IRAC PSF Photometry for Very Bright Stars}
\shortauthors{K. Su et al.}

\begin{document}

\title{Accurate Photometry of Saturated Stars Using the Point-Spread-Function Wing Technique with {\it Spitzer}}

\correspondingauthor{Kate Su}
\email{ksu@as.arizona.edu}

\author[0000-0002-3532-5580]{Kate Y.~L.~Su}
\affiliation{Steward Observatory, University of Arizona, 933 N Cherry Avenue, Tucson, AZ 85721-0065, USA}
\author[0000-0003-2303-6519]{G. H. Rieke}
\affiliation{Steward Observatory, University of Arizona, 933 N Cherry Avenue, Tucson, AZ 85721-0065, USA, also Department of Planetary Sciences}
\author[0000-0001-9910-9230 ]{M. Marengo}
\affiliation{Department of Physics and Astronomy, Iowa State University, 2323 Osborn Drive, Physics 0012, Ames, Iowa 50011-3160}
\author[0000-0001-8291-6490 ]{Everett Schlawin}
\affiliation{Steward Observatory, University of Arizona, 933 N Cherry Avenue, Tucson, AZ 85721-0065, USA}


\begin{abstract}
We report {\it Spitzer} 3.6 and 4.5 $\mu$m photometry of 11 bright stars relative to Sirius, exploiting the unique optical stability of the {\it Spitzer Space Telescope} point spread function (PSF). {\it Spitzer}'s extremely stable beryllium optics in its isothermal environment enables precise comparisons in the wings of the PSF from heavily saturated stars. These bright stars stand as the primary sample to improve stellar models, and to transfer the absolute flux calibration of bright standard stars to a sample of fainter standards useful for missions like {\it JWST} and for large groundbased telescopes. We demonstrate that better than 1\% relative photometry can be achieved using the PSF wing technique in the radial range of 20--100\arcsec\ for stars that are fainter than Sirius by 8 mag (from outside the saturated core to a large radius where a high signal-to-noise profile can still be obtained). We test our results by (1) comparing the [3.6]$-$[4.5] color with that expected between the WISE W1 and W2 bands, (2)  comparing with stars where there is accurate $K_{\text{S}}$ photometry, and (3) also comparing with relative fluxes obtained with the DIRBE instrument on COBE. These tests confirm that relative photometry is achieved to better than 1\%.  
\end{abstract}


\section{Introduction} \label{sec:intro}


Achieving high accuracy over huge dynamic range is a fundamental challenge in determining an accurate absolute calibration of astronomical photometry. Direct comparisons of stellar outputs with calibrated flux sources can generally only be made for very bright stars, but such stars are typically too bright to measure with the same instruments that are used for nearly all astronomical measurements. This challenge has been met in the optical making use of the behavior of the Hubble Space Telescope ({\it HST}) Space Telescope Imaging Spectrograph (STIS) Charge Coupled Device (CCD) detectors \citep{bohlin04}. There, the transfer from bright to faint stars depends on the recovery of virtually all of the charge bloomed by the saturated image of a bright star, with negligible losses into the channel stops of the CCD. Infrared detectors do not share this beneficial behavior, so a different strategy is needed for direct transfer over the required dynamic range. 

Fortunately, the superb optical stability of the Spitzer Space telescope ({\it Spitzer}) and the stable performance of the Infrared Array Camera (IRAC) provide an opportunity for direct transfers from very bright stars in the infrared. To do so, we have developed the technique pioneered by \citet{marengo09} to base the photometry on the wings of a PSF for heavily saturated images. We find that comparisons of stellar brightness can be made to the 1\% accuracy level with this approach. Our results are used in an accompanying paper (\citealt{rieke21}; hereafter Paper I) in an overall investigation of the transfer of the flux of Sirius to fainter stars. This transfer both is a step toward establishing Sirius as a new zero point standard (replacing Vega which has been shown to have multiple problems in this role) and to utilize the accurate absolute calibration of Sirius in photometry at faint levels. 

In addition to Sirius and HD 165459, the standard we use to help transfer from Sirius to much fainter stars, we have observed $\beta$ Gem (Pollux), a K0III star widely used for calibration, and $\alpha$ CMi (Procyon), selected as a very well studied star with an accurate interferometric diameter. We also selected eight additional stars of intermediate brightness ($\sim$3$^{rd}$ magnitude in the infrared) of spectral type mostly mid- to late-F, with accurate diameter measurements. It should be possible to model these stars very accurately to compare with our calibration results. With three exceptions, we have confined the study to dwarf stars of mid-F or earlier type since the later-type stars have CO absorption features in the IRAC 4.5 $\mu$m bandpass.

The paper is organized as follows. Section \ref{sec:reduction} summarizes the observations and basic reduction that lead to a high signal-to-noise (S/N) and well sampled stellar PSF outside the saturated core. Section \ref{sec:analysis} lays out the sequence of our analysis method and derives the fluxes of the targets relative to Sirius at both 3.6 and 4.5 $\mu$m. We then validate the accuracy of our results by a variety of methods in Section \ref{sec:check}: (1) comparing the observed color of [3.6]$-$[4.5] to the empirically derived, intrinsic color; (2) checking the consistency of our results with other very accurate photometry derived from measurements from multiple sources; and (3) also comparing with measurements from the DIRBE instrument on the COBE satellite. We give a short conclusion in Section \ref{sec:conclusion}. 

\begin{deluxetable}{lcccc}
\tablewidth{0pc}
\tablecaption{{\it Spitzer} Observations}
\label{tab:obs}
\tablehead{\colhead{Name} & \colhead{AOR} & \colhead{Observation} & \colhead{Frame Time} & \colhead{Dither}  \\
\colhead{}    & \colhead{}    & \colhead{Date}          & \colhead{(sec)}  & \colhead{Number} }
\startdata 
Sirius	    &   65280512	  & 2018-01-27    &    2    &       40   \\   
$\beta$ Gem    &   64890368	  & 2018-02-09    &    2    &       40   \\   
Procyon	    &   64891392	  & 2018-02-28    &    2    &       40   \\   
HD 30652    &   64892160	  & 2018-01-03    &    6    &      100   \\   
HD 102870   &   64892928	  & 2018-05-09    &    6    &      100   \\   
HD 142860   &   64892672	  & 2017-11-29    &    6    &      100   \\   
HD 19373    &   64891648	  & 2017-12-04    &    6    &      100   \\   
HD 126660   &   64892416	  & 2018-05-08    &    6    &      100   \\   
HD 215648   &   64893440	  & 2018-02-26    &    6    &      100   \\   
HD 173667   &   64893184	  & 2017-12-09    &    6    &      100   \\   
HD 10476    &   64891904	  & 2017-12-03    &    6    &      100   \\   
HD 165459   &   64893696	  & 2017-12-14    &   30    &      100    
\enddata
\end{deluxetable}

\begin{figure*}
    \includegraphics[width=\textwidth]{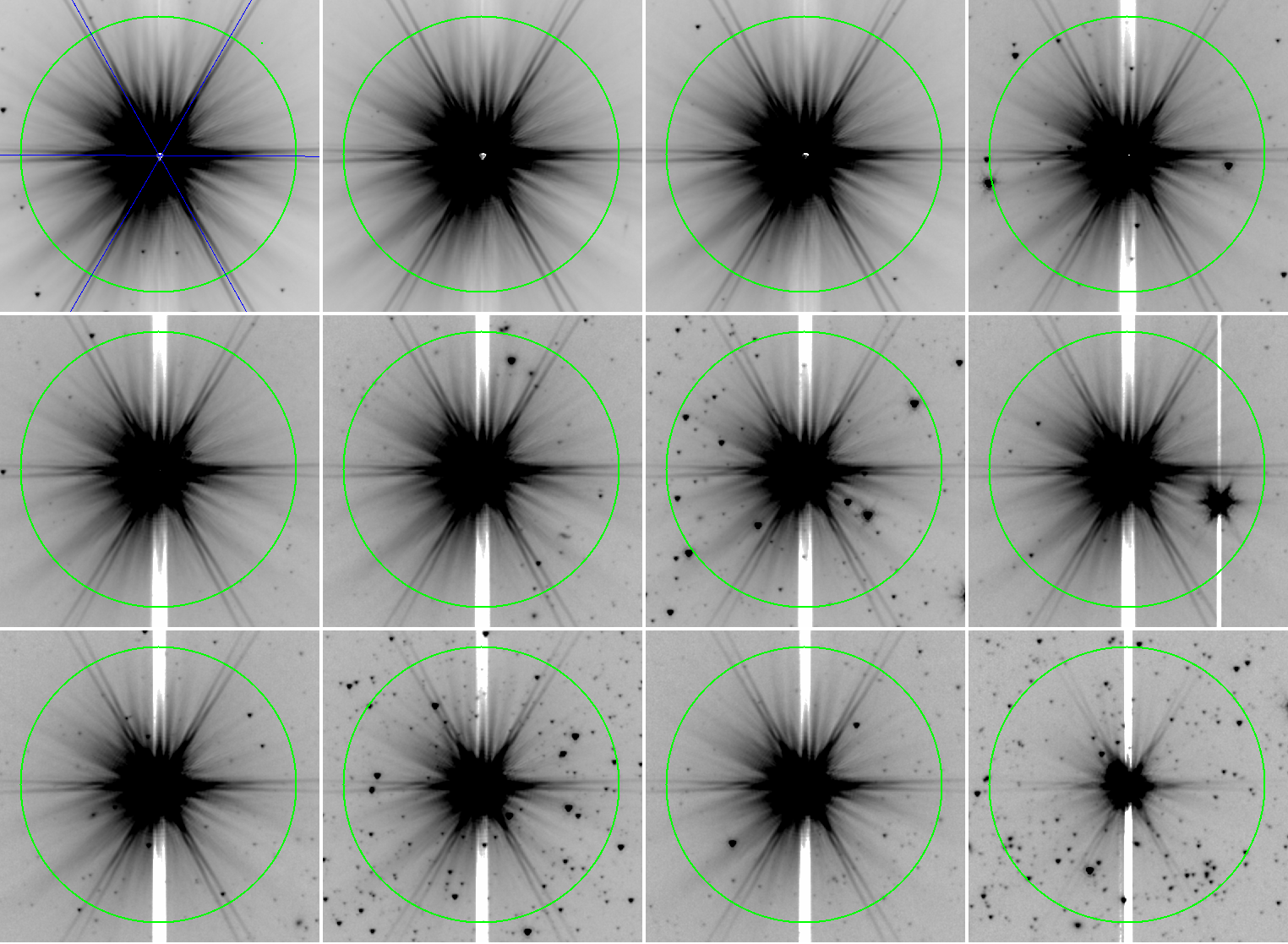}
    \caption{Final mosaics at 3.6 $\mu$m for the 12 stars in this study. The green circle in each panel marks a radius of 100\arcsec\ circle centered at the star. The stars from left to right and top to bottom are: Sirius, $\beta$ Gem, Procyon, HD 30652, HD 102870, HD 142860, HD 19373, HD 126660, HD 215648, HD 173667, HD 10476, HD 165459. The intercept of the three blue lines in the Sirius data (top left panel) marks the best-guess centroid position using the diffraction spikes. }
    \label{fig:finalmosaics}
\end{figure*}

\section{Observation and Data Reduction} \label{sec:reduction}

The observations presented here were taken in 2017/2018 during the {\it Spitzer} warm mission (PID 13211, PI: Rieke) as part of the effort to advance absolute calibration for JWST and future applications. Twelve bright stars were observed with 
IRAC channel 1 and 2 (I1 \& I2) using the full-array mode. To ensure good sampling of the PSF and minimize the effects of intrapixel sensitivity variations of the detector, all observations were taken with 40 or 100 dithers using a large-scale cycling mode. The frame time was chosen based on the brightness of the star to ensure high S/N data were obtained around the saturated core of the star. Details regarding the observations are given in Table \ref{tab:obs}.  

All data were first processed by the {\it Spitzer} Science Center using the pipeline version S19.2.0 for basic reduction of all the individual frames (Basic Calibration Data, BCD). To better sample the wings of the PSF, we then generated the final combined mosaics using {\it mopex} with subsampling by a factor of 5, i.e., the resultant platescale is 0\farcs24 per resampled pixel. We input all BCD files along with associated uncertainty and mask files and used all the default parameters in {\it mopex} when generating combined mosaics (except for the subsample factor). The final combined 3.6 $\mu$m images are shown in Figure \ref{fig:finalmosaics}. The saturated core (the region with radius, $r$, from the center $\lesssim$10\arcsec) and the area of the data affected by the ``pull-down" effect (a vertical strip across the saturated star) were both masked out during the further analysis.

\begin{figure*}
    \centering
    \includegraphics[width=0.48\linewidth]{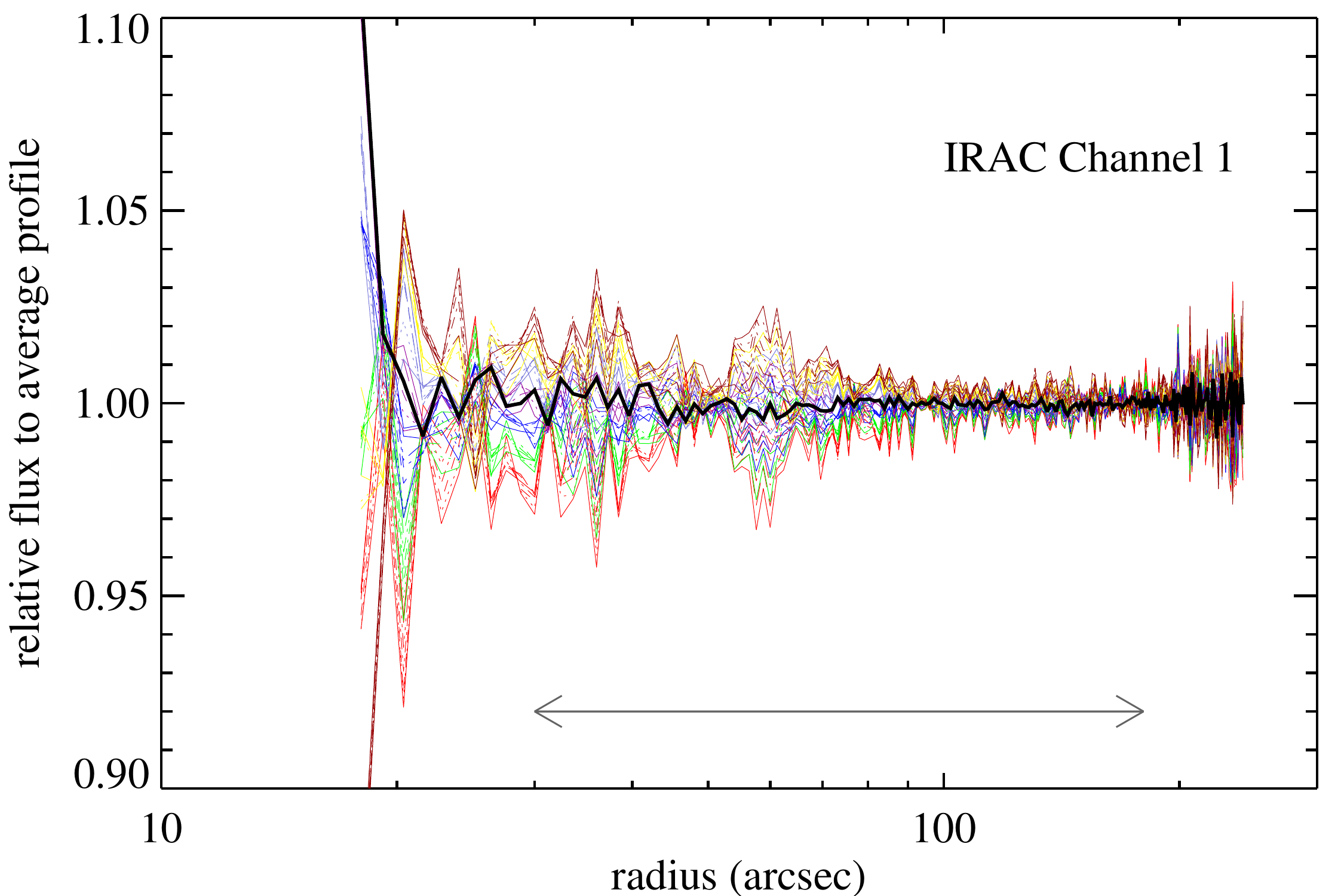}
    \includegraphics[width=0.48\linewidth]{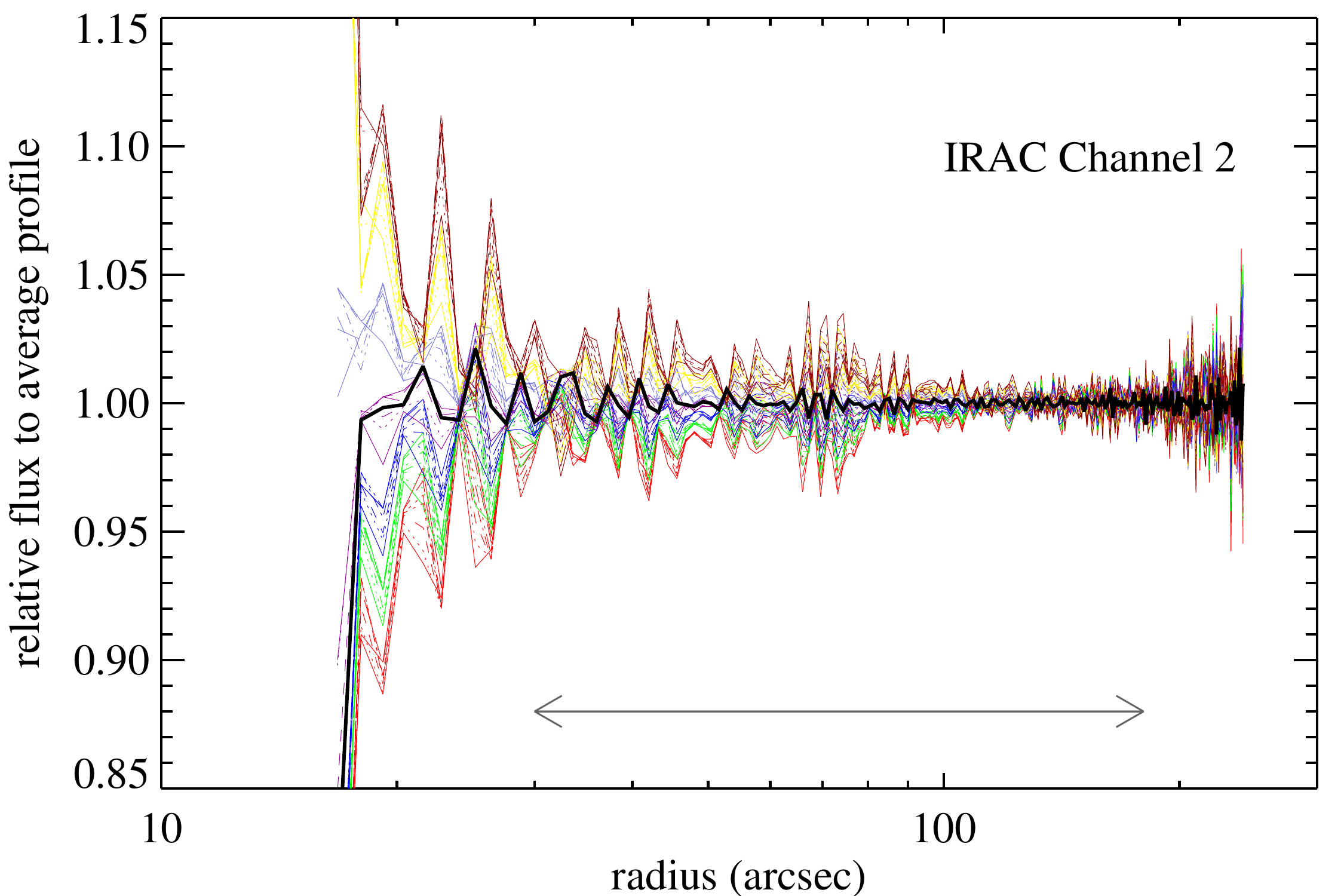}
    \caption{Relative radial profiles using 49 different center positions for Sirius (thin colored lines) with the left panel for 3.6 $\mu$m data and right for 4.5 $\mu$m. In each panel, the thick black line is the profile that has the minimum deviation from the average profile within the range of 30\arcsec--180\arcsec\ marked by the gray arrow.}
    \label{fig:sirius_prof_ratio}
\end{figure*}

\begin{figure*}
    \includegraphics[width=\textwidth]{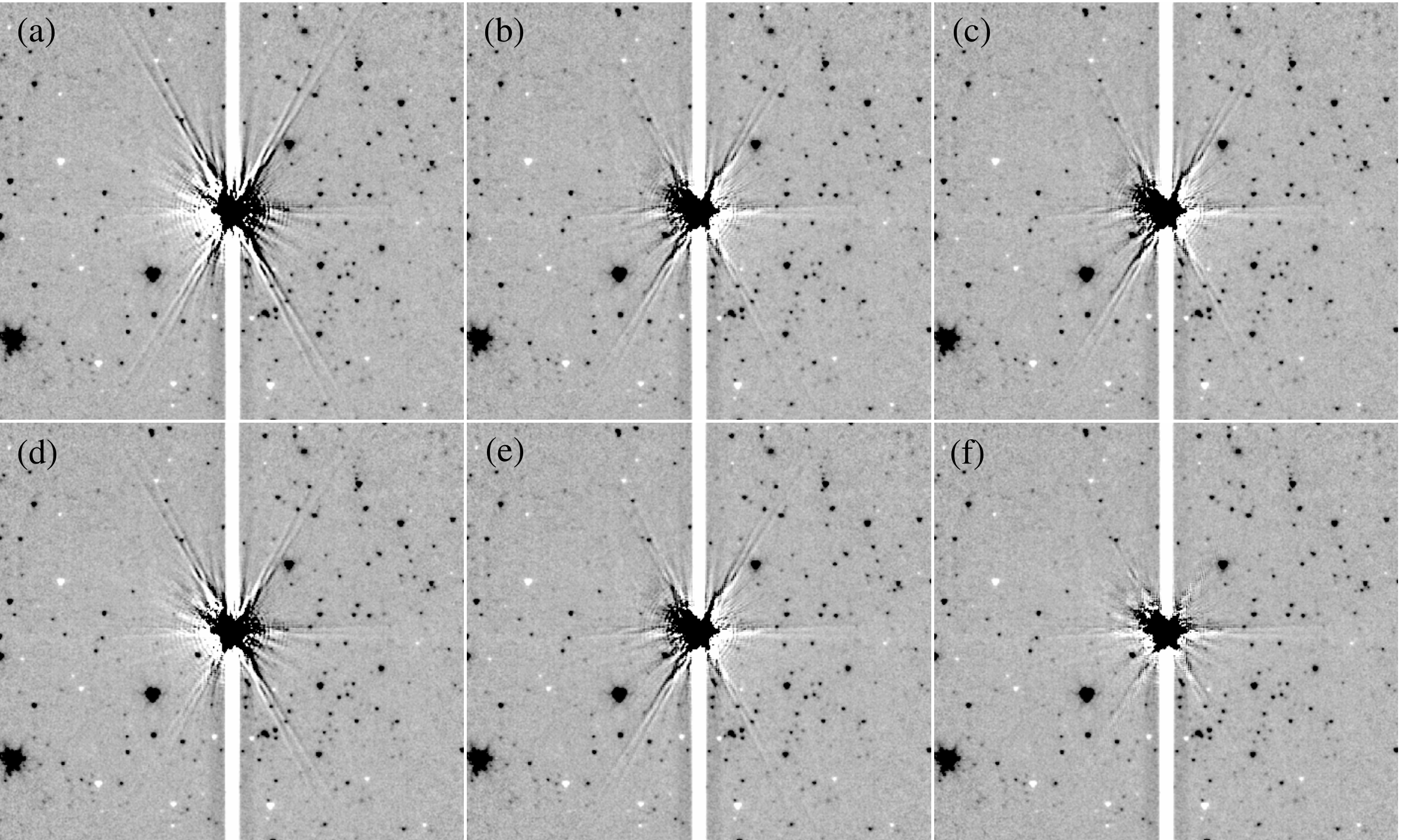}
    \caption{Example of the difference images between HD 10476 and Sirius at 6 different shift positions from (a) to (e) as the shift position gets closer to the best registered position (f) where the prominent diffraction spikes disappear.}
    \label{fig:diff_example}
\end{figure*}

\section{Analysis and Results} \label{sec:analysis}

\subsection{Determining the Centroid for Sirius}

Our technique requires accurate centering of the images of the reference star (Sirius) and the star of interest. The PSF of IRAC is roughly point- and axi-symmetric; that is, its center position can be determined by examining the non-saturated flux distribution around the star (i.e., radial profile). Determining the accurate position of the centroid of a heavily saturated star image is a two-step process. We first determine a ``best-guess" center using the PSF diffraction spikes, and then the centroid is fine-tuned using the non-saturated radial profile distribution of the star.  The ``best-guess" center is determined by the intercepts of the diffraction spikes by examination of the image. In image display software (such as $\it ds9$), an initial guess of the centroid can be determined by drawing a symmetric line for each pair of diffraction spikes diagonally and horizontally (see the top left panel in Figure \ref{fig:finalmosaics} for an example). These lines intercept within a few resampled pixels if they are truly symmetric for a pair of the spikes, and a rough center can be determined as the center of the intercepted area. Using this rough center, a radial profile is then constructed using the median value within a concentric annulus that has a width of 5 resampled pixels (1\farcs2). Multiple radial profiles were obtained by shifting the center by $\pm$3 resampled pixels (i.e., a search area of 1\farcs7$\times$1\farcs7, roughly the full-width at half maximum of the PSF)  around the best-guess center in both $x$ and $y$ directions, resulting in a total of 7$\times$7=49 profiles. Comparing all these radial profiles, we found that outside 30\arcsec\ all the profiles agree well, within a few \%, while the interior profiles between 10\arcsec--20\arcsec\ depart from each other but  still agree within 10\%. A fiducial radial profile is computed by averaging all the profiles. Figure \ref{fig:sirius_prof_ratio} shows an example of all the profiles normalized to the fiducial one. The best centroid was determined from the profile that has the minimum deviation in the $\chi^2$ sense from the average profile between 30\arcsec\ and 180\arcsec. The profile using this best centroid is also shown as the thick, black line in Figure \ref{fig:sirius_prof_ratio}. At both bands, the shifts between the initial guess and the best centroid are small, within $\pm$1 resampled pixel. We note that the exact centroid position has no impact on the radial profile outside the saturated core ($r>$10\arcsec), and our technique (as described below) is not sensitive to this ``best-guess" center as long as we can perfectly align two saturated images.

\begin{figure*}
    \includegraphics[width=0.48\textwidth]{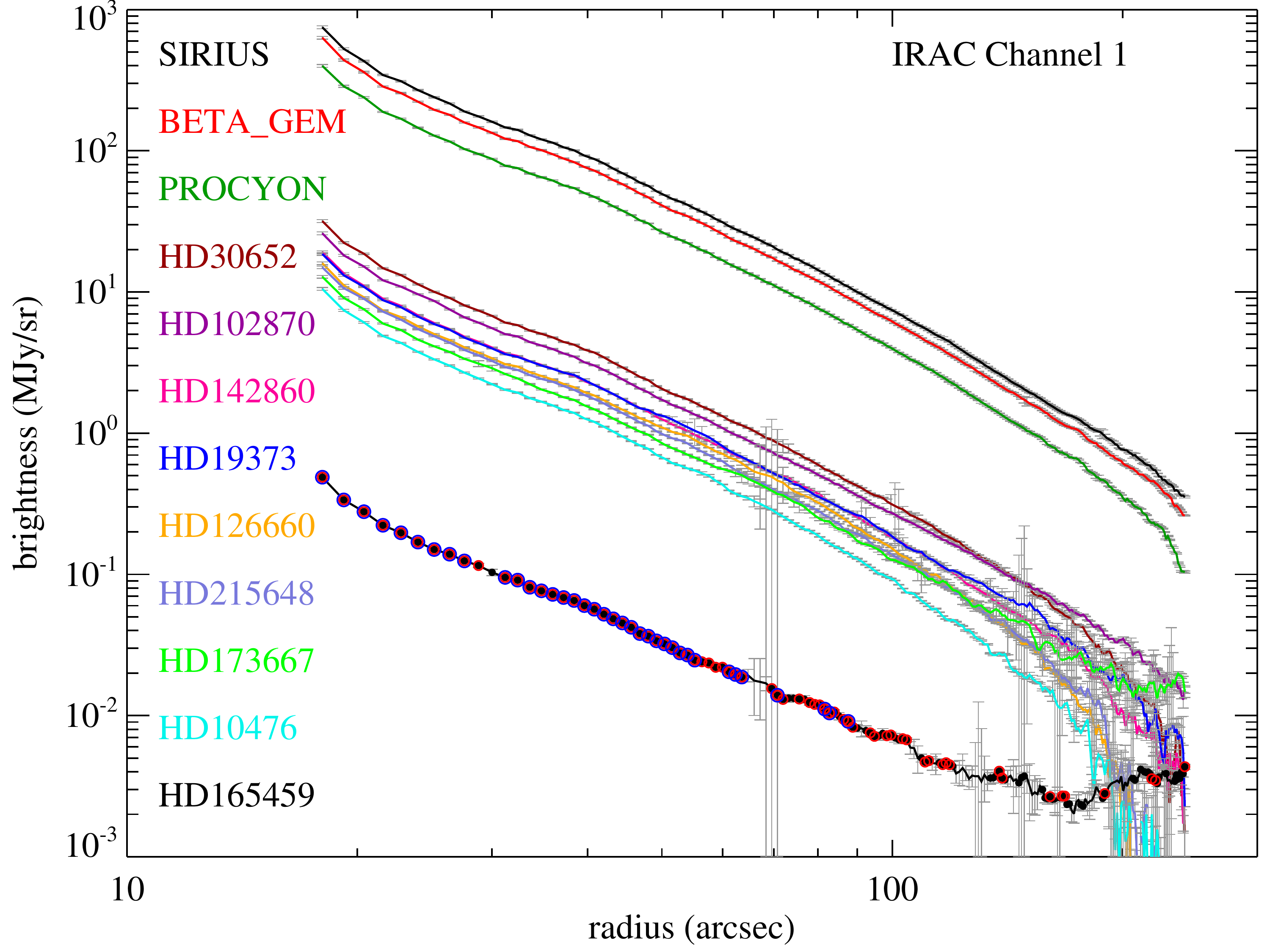}
    \includegraphics[width=0.48\textwidth]{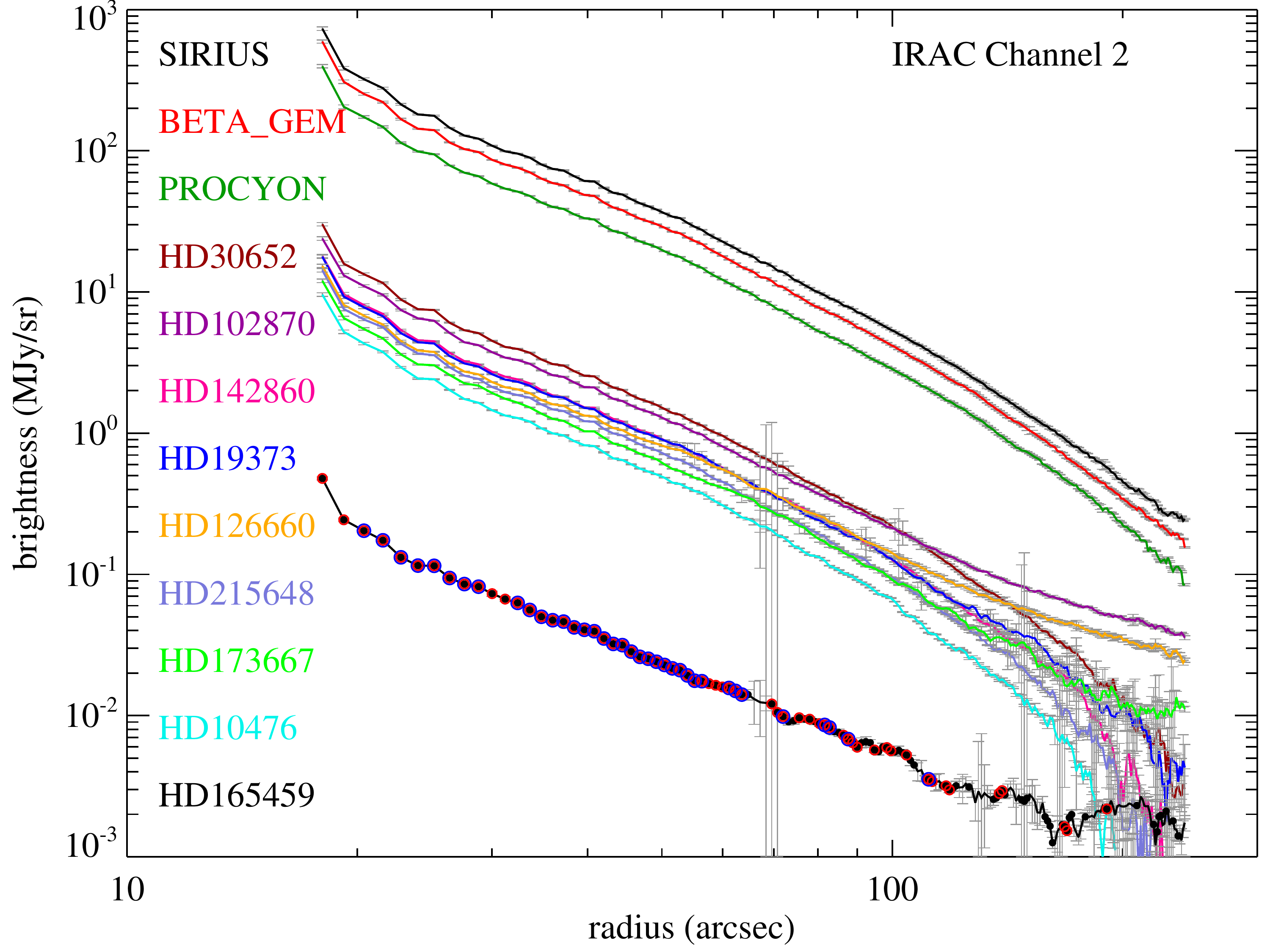}
    \caption{Radial profiles of all targets in the 3.6 $\mu$m (left panel) and 4.5 $\mu$m (right panel) bands, respectively. Different stars are plotted with different colors given on the plot. All radial profiles are self similar within 100\arcsec\ (a dynamical range of $\gtrsim$100 per star) while the profiles outside that range depart due to small background offsets. For the faintest target HD165459, we also use various symbols to illustrate the signal-to-noise (S/N) ratio: solid black dots represent points with S/N$\geq$10, red circles for S/N$\geq$20, and blue circles for S/N$\geq$40.}
    \label{fig:radcomp}
\end{figure*}

\subsection{Aligning Images of the Star of Interest and Sirius}

Instead of determining an accurate absolute centroid for the star of interest, we only care about the centroid position relative to that of Sirius (i.e., the goal is to align the two images perfectly). Achieving this alignment is also a two-step process. First, a best-guess center of the stellar image is determined using the symmetry of the diffraction spikes, similarly to the Sirius data as presented in the previous section. Second, the best alignment is determined by minimizing the residuals in difference images between Sirius and the target of interest. Because of the flux difference between the star and Sirius, the images are first scaled to match in brightness. The scale factor is determined by matching the surface brightness level in the radial profiles at a radius of 40\arcsec. The scaled target image is roughly centered on the centroid of the Sirius image, and then 11$\times$11 shifted images are generated over a range of $\pm$5 resampled pixels around the ``best-guess" centroid. Difference images are generated by subtracting the scaled and shifted images from the Sirius image. The best shift position is then determined visually by examining the difference images (see Figure \ref{fig:diff_example} for an illustration). When the two images are mis-aligned, the PSF structures (spikes and diffraction rings) in the difference image are amplified as echos in a positive and negative pattern, but these artifacts mostly disappear when the two images are perfectly aligned. For each of the stars, the best match (shift values in $x$ and $y$ directions) was determined independently for the data from both channels 1 and 2. 

All 11 stars were measured in both bands relative to Sirius using this procedure.  Once a star was aligned with Sirius, its final radial profile was constructed using the best centroid position, as shown in Figure \ref{fig:radcomp}. To construct the final radial profiles, we used the median value within the concentric annulus as the radial flux value, and determined its associated uncertainty as the standard deviation in the annulus divided by the square root of the number of resampled pixels within the annulus. Using the median value ensures that the profile is not biased by bright field stars. As shown in Figure \ref{fig:finalmosaics}, there is a bright (saturated) field star in the field of HD 126660 roughly at a radius of 70\arcsec, resulting in larger uncertainties near that radius. The profiles between HD 126660 and HD 215648 are very similar within 40\arcsec, but depart slightly due to the saturated field star in the HD 126660 image and the different values of the background offset. Nevertheless, all radial profiles are self similar in each band up to a radius of 120\arcsec\ from the star. Outside that radius, the radial profiles depart from that of Sirius, and more severely for fainter stars where any small background offset affects the profile strongly. We note that the background value is not the true background value in the field, rather it is a relatively small background mis-match arising when combining the mosaics; therefore, it can be either positive or negative.  Overall, the final derived radial profiles have good S/N over a dynamical range of 10$^4$ for bright stars like Sirius, $\beta$ Gem and Procyon, $\sim$10$^2$ for the faintest star, HD 165459, and 10$^3$ for the rest of the targets.

\begin{deluxetable*}{lcccccccccccc}
\tablewidth{0pc}
\tablecaption{Derive Flux Ratio Relative to Sirius at Three S/N Thresholds Using Method 1 \label{tab:fluxratio}}
\tablehead{
\colhead{}  & \multicolumn{4}{c}{S/N=20} & \multicolumn{4}{c}{S/N=30} & \multicolumn{4}{c}{S/N=40} \\ 
\colhead{Name} &\colhead{N} &\colhead{$\bar{f}$} &\colhead{$\tilde{f}$}&\colhead{$\sigma_f$} & \colhead{N} &\colhead{$\bar{f}$} &\colhead{$\tilde{f}$}&\colhead{$\sigma_f$} &\colhead{N} &\colhead{$\bar{f}$} &\colhead{$\tilde{f}$}&\colhead{$\sigma_f$}}
\startdata
 \multicolumn{13}{l}{Channel 1} \\
$\beta$ Gem     & 87 & 1.2040 & 1.2044 & 0.0058 & 87 & 1.2040 & 1.2044 & 0.0058 & 87 & 1.2040 & 1.2044 & 0.0058 \\
Procyon         & 73 & 1.8563 & 1.8576 & 0.0120 & 73 & 1.8563 & 1.8576 & 0.0120 & 73 & 1.8563 & 1.8576 & 0.0120 \\
HD 30652         & 71 & 23.737 & 23.738 & 0.188  & 68 & 23.745 & 23.739 & 0.186  & 67 & 23.749 & 23.739 & 0.185 \\
HD 102870        & 38 & 28.760 & 28.775 & 0.159  & 38 & 28.760 & 28.775 & 0.159  & 38 & 28.760 & 28.775 & 0.159 \\
HD 142860        & 42 & 39.504 & 39.542 & 0.306  & 42 & 39.504 & 39.542 & 0.306  & 42 & 39.504 & 39.542 & 0.306 \\
HD 19373         & 38 & 39.686 & 39.680 & 0.451  & 37 & 39.706 & 39.680 & 0.440  & 36 & 39.704 & 39.680 & 0.446 \\
HD 126660        & 27 & 46.939 & 47.257 & 0.807  & 26 & 47.023 & 47.258 & 0.691  & 25 & 47.103 & 47.258 & 0.572 \\
HD 215648        & 27 & 49.541 & 49.549 & 0.476  & 27 & 49.541 & 49.549 & 0.476  & 27 & 49.541 & 49.549 & 0.476 \\
HD 173667        & 37 & 57.637 & 57.861 & 0.920  & 31 & 57.867 & 57.886 & 0.694  & 29 & 57.939 & 57.938 & 0.658 \\
HD 10476         & 26 & 72.061 & 72.187 & 0.774  & 26 & 72.061 & 72.187 & 0.774  & 25 & 72.014 & 72.175 & 0.751 \\
HD 165459        & 19 & 1552.60 & 1544.38 & 18.11& 18 & 1554.71 & 1552.71 & 16.06& 17 & 1555.66& 1552.71 & 16.02 \\ 
\hline
 \multicolumn{13}{l}{Channel 2} \\
$\beta$ Gem      & 62 & 1.2720 & 1.2721 & 0.0142 & 62 & 1.2720 & 1.2721 & 0.0142 & 61 & 1.2724 & 1.2721 & 0.0141 \\
Procyon          & 77 & 1.8646 & 1.8661 & 0.0162 & 77 & 1.8646 & 1.8661 & 0.0162 & 75 & 1.8645 & 1.8661 & 0.0154 \\
HD 30652          & 43 & 24.107 & 24.090 & 1.409  & 42 & 24.101 & 24.090 & 1.376  & 38 & 24.08  & 24.09 & 1.43 \\
HD 102870         & 29 & 28.796 & 28.851 & 0.435  & 29 & 28.796 & 28.851 & 0.435  & 27 & 28.74  & 28.77 & 0.49 \\
HD 142860         & 34 & 39.850 & 39.964 & 0.507  & 34 & 39.850 & 39.964 & 0.507  & 29 & 39.89  & 39.97 & 0.49 \\
HD 19373          & 51 & 41.085 & 41.096 & 0.366  & 48 & 41.119 & 41.130 & 0.357  & 45 & 41.12  & 41.11 & 0.42 \\
HD 126660         & 23 & 46.736 & 46.666 & 2.838  & 23 & 46.736 & 46.666 & 2.625  & 21 & 46.68  & 46.64 & 2.11 \\
HD 215648         & 36 & 50.506 & 50.612 & 0.585  & 36 & 50.506 & 50.612 & 0.585  & 34 & 50.50  & 50.61 & 0.78 \\
HD 173667         & 21 & 58.033 & 58.417 & 1.895  & 19 & 58.256 & 58.616 & 1.643  & 12 & 58.56  & 58.88 & 1.40 \\
HD 10476          & 28 & 74.355 & 74.543 & 0.933  & 26 & 74.356 & 74.543 & 0.922  & 24 & 74.39  & 74.56 & 0.90 \\
HD 165459         & 18 & 1556.25 & 1565.47 & 47.90& 17 & 1566.77 & 1566.19 & 35.19& 15 & 1565.34& 1565.47 & 33.49 
\enddata
\tablecomments{N is the number of points used to derive the average ($\bar{f}$), median ($\tilde{f}$) and standard deviation ($\sigma_{f}$) of the flux ratio.}
\end{deluxetable*}

\begin{figure*}
    \figurenum{5a}
    \includegraphics[width=\textwidth]{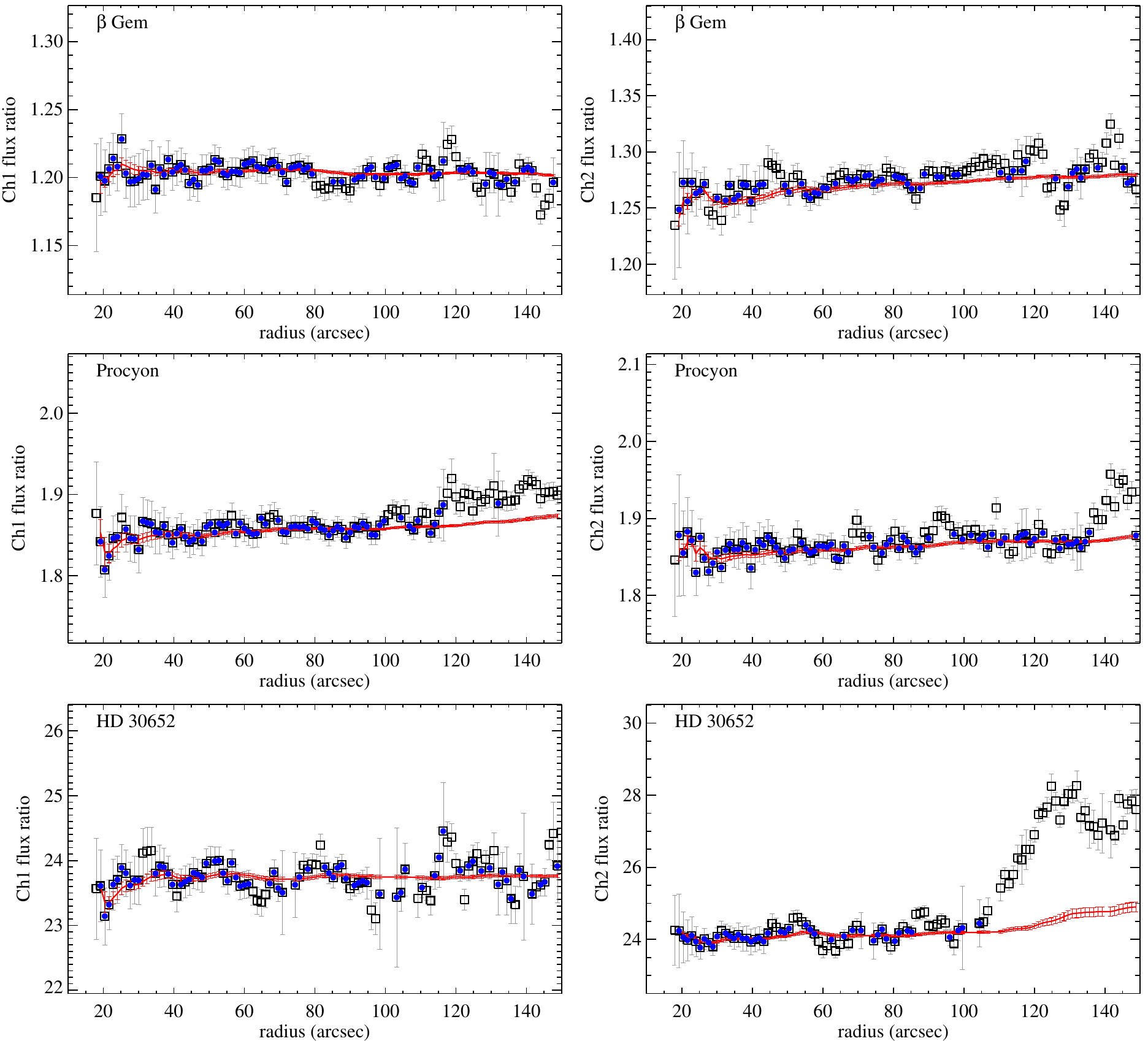}
    \caption{Radial flux ratio between Sirius and the star of interest with the name given on the upper left corner of each panel. For each row, the left panel is for channel 1 and the right panel for channel 2. Data points are shown as open, black squares with 1 $\sigma$ error bars, and the accumulative weighted average is shown with a solid red line. Blue dots show the data points that are used to compute the final flux ratio (i.e., having S/N greater than the cutoff threshold (20 in this case) and within $\pm$1 $\sigma$ of the average profile). }
    \label{fig:sn20}
\end{figure*}
\begin{figure*}
    \figurenum{5b}
    \epsscale{1.1}
    \plotone{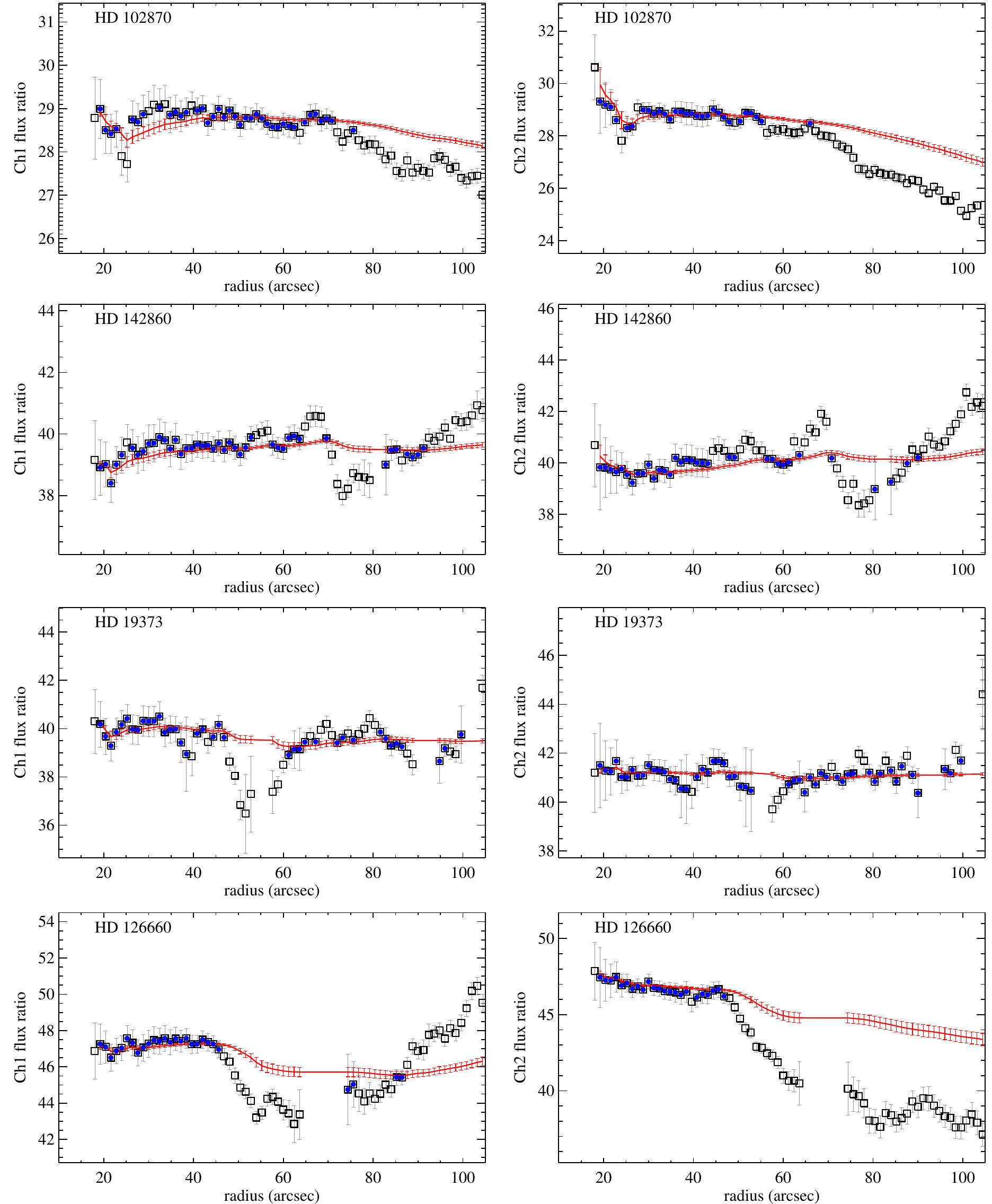}
    \caption{Similar to Figure \ref{fig:sn20} but for four additional stars. Note that data points that have S/N lower than the threshold (20 in this case) are not shown. } 
    \label{fig:sn20b}
\end{figure*}
\begin{figure*}
    \figurenum{5c}
    \epsscale{1.1}
    \plotone{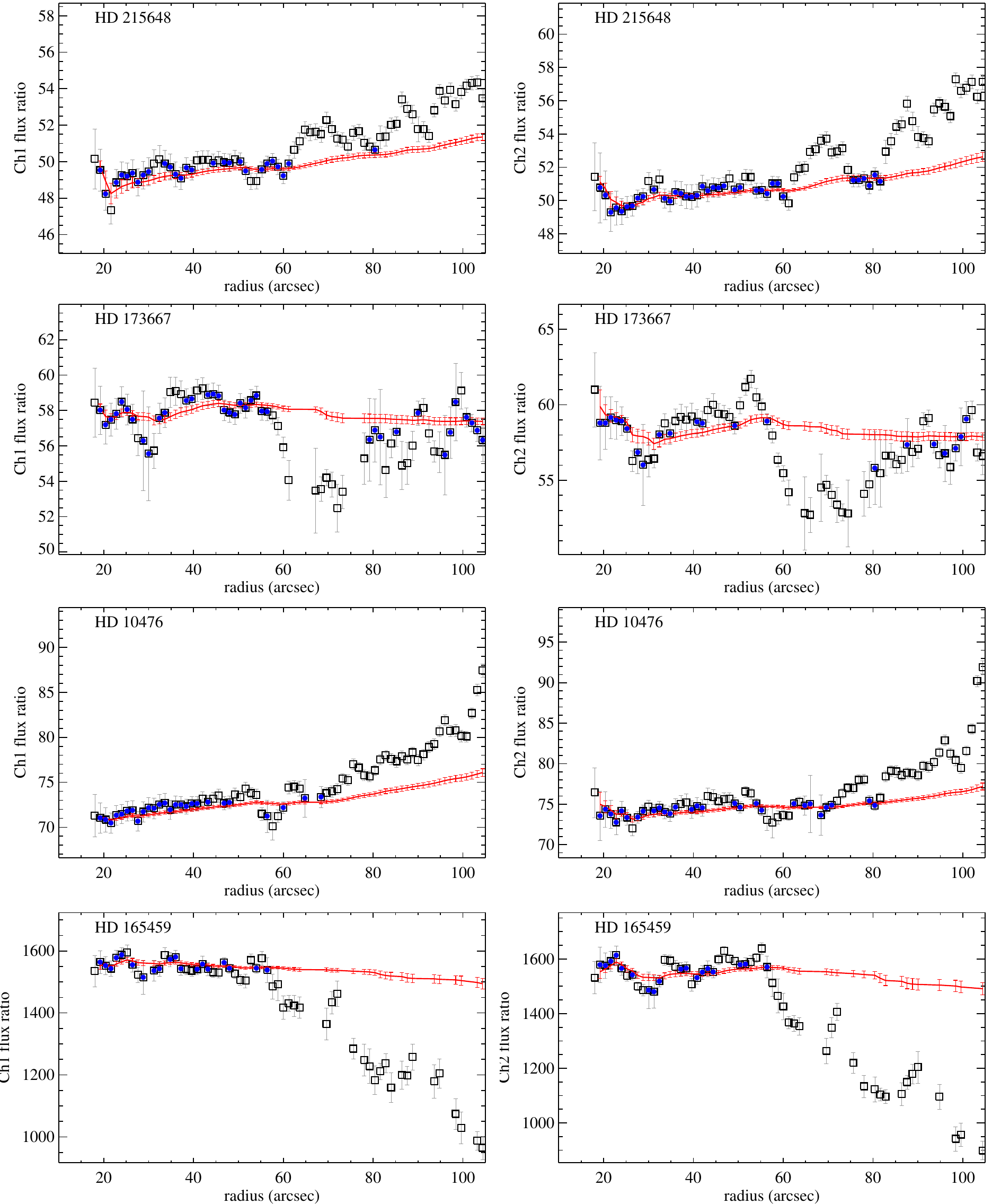}
    \caption{Similar to Figure \ref{fig:sn20} but for the rest of the stars} 
    \label{fig:sn20c}
\end{figure*}

\subsection{Determining the Relative Brightness} \label{subsec:radprofcom}

We determine the relative brightness between the star of interest and Sirius by comparing their radial profiles. The brightness difference between a star and Sirius should be the flux ratio of the two profiles radius by radius, which is expected to be constant in an ideal situation (e.g., PSF is stable and the median radial profile is a true representative of the PSF structure, with no offset from the background). In reality, there are a number of challenges in comparing the radial profiles. Bright background sources within the extended PSFs and their associated instrumental artifacts could bias the results; we masked obvious examples. However, subtle changes in the underlying background are harder to detect. In fact, by necessity the background level is determined far from the stellar positions, while the radial profile comparison is dominated by closer-in positions, so offsets are definitely possible. Other sources of uncertainty can arise, resulting in inconsistencies in the profiles indicated by elevated noise in their values at some radii. 

Figure 5 shows the radial distribution of the flux ratio between Sirius and each target star. For bright stars like $\beta$ Gem and Procyon, the flux ratio profile is relatively flat within the uncertainties (computed by the error propagation of the two profiles) within radii of 120\arcsec. For fainter stars as shown in Figure \ref{fig:sn20b} and \ref{fig:sn20c}, various non-flat structures are seen, which are not necessarily the same for both bands. We adopted two different methods to compute the average values of the flux ratios.

\underline{Method 1\ } {\it Rejecting Points Using the Accumulative Average Profile}.  As mentioned earlier, in an ideal case we expect a flat radial distribution in the flux ratio. This is certainly the case at small radii as the brightness of the star drops radially. Therefore, the flux ratio at small radii should carry more weight in determining the true flux ratio. The situation is similar to determining the optimal aperture size in photometry: we need to find a size that is large enough to contain most of the signal but small enough so that artifacts from the background placement do not dominate the signal. For each of the profiles, we therefore computed the accumulative average profile (in flux ratio) such that at each radial point the value is the weighted average of all the data points prior to that radius (similar to the enclosed circle energy distribution). Because the accumulative average is more weighted by the values in the previous radii, the average is less affected by adding a new deviant point and remains relatively flat for bright stars (see Figure \ref{fig:sn20}). We then used the accumulative average profile as the guide and computed the average flux ratio by retaining the points that are within $\pm$1 $\sigma$ of the accumulative average profile (blue dots in Figure 5). To reserve only the high S/N points, a threshold was also imposed when computing the accumulative average profile. The higher the threshold value, the smaller the number of points that are qualified to derive the average value. We have tried several different thresholds as given in Table \ref{tab:fluxratio}. For bright sources, this threshold does not make any difference; furthermore, the average and median flux ratios are also indistinguishable within 1 $\sigma$ uncertainty.

\underline{Method 2\ } {\it Extending the Fitting Range by Adjusting the Small Background Offset}. As shown in Table \ref{tab:fluxratio}, the number of points used to derive the flux ratio using Method 1 is in the range of a few dozens for fainter sources. The situation can be improved if one allows for a small adjustment in the background offset in the radial profile. Conservatively, we first rejected all points that have S/N less than 40 for the target star (there are no such regions for Sirius). We adjusted the constant added to the target profile to make the ratio independent of radius (i.e., flat) so far as possible, giving most weight to the points at smaller radii. We trimmed the sample of points at either end of the profile that deviated more than any others from the fit, and carried out this process iteratively until such extreme points were not included. This step allowed a systematic way to remove points where various systematic effects resulted in a departure from the true radial profile of the PSF. Finally, we took the ratio as the median of the accepted points (the mean did not differ significantly, but the median is relatively insensitive to outliers). This procedure was used for both bands independently.

\begin{table*}
\tablewidth{0pc}
\begin{center}
\caption{Derived Photometry Relative to Sirius \label{tab:finalresults}}
\begin{tabular}{llccccccccccr}
\hline
\hline
 & & & & \multicolumn{6}{c}{PSF Photometry relative to Sirius\tablenotemark{a}} & & \multicolumn{2}{c}{relative to Sirius}   \\
 \cline{5-10} \cline{12-13}
Name & Sp Type & $V$ & $K_{\text{S}}$ & \multicolumn{3}{c}{[3.6]} & \multicolumn{3}{c}{[4.5]} & measured  & predicted & residual   \\
 & & & & M1 & M2  & Avg  & M1 & M2  & Avg & [3.6]$-$[4.5] & [3.6] & [3.6] \\
\hline
$\beta$ Gem	&K0III	&	1.14	&	-1.118	&	0.202	&	0.202	&	0.202	&	0.262	&	0.260	&	0.261	&	-0.059	&	0.207\tablenotemark{b}	&	0.005   \\
Procyon	    &F5IV-V 	&	0.37	&	-0.677	&	0.672	&	0.669	&	0.671	&	0.676	&	0.673	&	0.675	&	-0.004	&	0.689	&	0.018   \\
HD  30652	&F6IV-V	&	3.19	&	2.076	&	3.439	&	3.441	&	3.440	&	3.454	&	3.444	&	3.449	&	-0.009	&	3.441	&	0.001  \\
HD 102870	&F8.5IV-V&	3.60	&	2.293	&	3.647	&	3.656	&	3.652	&	3.646	&	3.651	&	3.649	&	0.003	&	3.656	&	0.004   \\
HD 142860	&F6V	&	3.84	&	2.623	&	3.992	&	3.989	&	3.991	&	4.002	&	3.992	&	3.997	&	-0.006	&	3.987	&	-0.004  \\
HD  19373	&G0IV-V	&	4.05	&	2.650	&	3.997	&	4.004	&	4.001	&	4.035	&	4.029	&	4.032	&	-0.032	&	4.012	&	0.011   \\
HD 126660	&F7V	&	4.05	&	2.808	&	4.183	&	4.184	&	4.184	&	4.173	&	4.188	&	4.181	&	0.003	&	4.172	&	-0.012  \\
HD 215648	&F6V	&	4.20	&	2.900\tablenotemark{c}	&	4.237	&	4.229	&	4.233	&	4.258	&	4.240	&	4.249	&	-0.016	&	4.244	&	0.011   \\
HD 173667	&F5.5IV-V&	4.20	&	3.039	&	4.407	&	4.413	&	4.410	&	4.419	&	4.405	&	4.412	&	-0.002	&	4.404	&	-0.006  \\
HD  10476	&K0/K1V	&	5.24	&	3.280	&	4.644	&	4.637	&	4.641	&	4.679	&	4.666	&	4.673	&	-0.032	&	4.634	&	-0.007  \\
HD 165459	&A3V	&	6.867
&	6.587	&	7.980	&	7.984	&	7.982	&	7.987	&	7.978	&	7.983	&	0.000	&	7.969	&	-0.013  \\
\hline
\end{tabular}
\end{center}
\tablenotetext{a}{Further discussion of these stars can be found in Paper I. All the photometry is in units of magnitudes as derived from Method 1 (M1), Method 2 (M2) and straight average (Avg). We use the relation between $K_{\text{S}}$ and IRAC Band 1 from equation (1) to predict the [3.6] magnitude from that at $K_{\text{S}}$. We adopt $K_{\text{S}} = [3.6] =-$1.395 mag for Sirius \citep{rieke21} for the values listed in the last two columns.}
\tablenotetext{b}{$\beta$ Gem is an evolved star that is not suitable for the color relation established for dwarfs; the expected value is derived as described in Paper 1.}
\tablenotetext{c} {This star has a close companion; the effect of the companion has been removed from the stellar magnitude, but it must be added in (0.019 mag) for the transfer to Sirius.}
\end{table*}

The flux ratio so determined can be converted to a magnitude relative to Sirius as 2.5~log($f$) where $f$ is the flux ratio. The standard deviation derived in Method 1 yields an uncertainty of 0.002--0.008 mag at [3.6], and 0.005--0.02 mag at [4.5]. The derived flux ratios using the two independent methods agree very well. Table \ref{tab:finalresults} lists the relative photometry of the targets using the two methods in units of magnitudes at both bands (we adopted the average value in Method 1 with S/N=40 threshold for direct comparison with the Method 2 value). Given the agreement, we simply average the results from the two methods and use the averages for further analysis. Our results suggest that the PSF wing technique is capable of delivering relative photometry as accurate as $\sim$1\% level at both bands.

\section{Tests of the Photometry} 
\label{sec:tests}

\subsection{Comparison with Other Sources of Photometry}
\label{sec:check}

We now test the approach to gain confidence that we can use our 3.6 and 4.5 $\mu$m photometry as one of the steps to establish absolute flux calibration between Sirius and fainter stellar standards. We have used several approaches to check the accuracy of our photometry. The first one is to use the color information for stars with very similar spectral types. Our measurements in the two IRAC bands are independent: they are made with two separate optical trains and detectors and, of course, reduced separately. Thus, the uncertainty in the color difference between these bands reflects the photometric errors well.  There are 5 stars that have the same spectral type (F5/6 V) as shown in Table \ref{tab:finalresults}. Of them, HD 215648 has a M dwarf companion 11\farcs1 away \citep{raghavan10} within the saturated PSF core, making the measured color more negative in relative to Sirius than what it should be. We have assigned an approximate spectral type to this companion of M3V based on its $K_{\text{S}}$ mag and estimated that its effect on the [3.6]$-$[4.5] color is 0.003 mag. Making this adjustment, the average [3.6]$-$[4.5] color for the five stars is $-$0.0080$\pm$0.0034 mag, i.e., the rms scatter is 0.0067 mag. Another estimate of the expected color can be derived using synthetic photometry from the theoretical stellar models (BOSZ database of a F5V star and the corresponding CALSPEC spectrum of Sirius as in \citealt{bohlin2017,bosz}). The synthetic color is $-$0.0010, consistent with our measurements.  We can compare this with the tabulation of the similar WISE [W1]$-$[W2] color by \citet{pecaut13}, where we force zero color for A0V stars by subtracting the value for them. Averaging the value for F5V and F6V, we find a color of $-$0.0095. \citet{pecaut13} estimate the errors in the WISE colors to be $\sim$1\%.  These results suggest that our derived photometry is probably accurate to within $\sim$0.006 mag per band, consistent with the uncertainty derived in the last section.

The second method of checking our result makes use of the $V-K_{\text{S}}$ vs. $K_{\text{S}}-$[3.6] color-color relation. We adopt the relation from Paper I where it is derived  using high-quality $V$, $K_{\text{S}}$ and {\it Spitzer} 3.6 $\mu$m photometry as: 
\begin{equation}
\label{eqn:cc}
\begin{split}
    K_{\text{S}} - [3.6] = -0.001228\ x^4 +0.014382\ x^3 \\ 
    - 0.042158 x^2  + 0.057587 x,
\end{split} 
\end{equation}
\noindent
where $x=V-K_{\text{S}}$. This equation is a fit to 950 stars, each with photometric errors of only $\sim$2\%, so nominally it has small errors in its mid-range where we take the colors for this discussion. For the F-type (and the G0V) stars, the nominal errors are $<$0.002 mag. The dominant error may be due to the uncertainty in reddening, estimated in Paper I as only about 0.002 mag. In any case, the errors are small compared with 1\%. 

Using this relationship, we can then predict a star's [3.6] magnitude relative to Sirius. Note that this color relationship only applies to main-sequence stars and adopts a calibration system that references to Sirius (A0V star with $V= K_{\text{S}} = -$1.395 mag; Paper I). Since these bright stars are heavily saturated in 2MASS, the $K_\text{S}$ magnitudes listed in Table \ref{tab:finalresults} are derived and transferred from heritage infrared photometry by Paper I\footnote{The $K_{\text{S}}$ photometry of HD 10476 is just from a single measurement by \citet{alonso94} and hence of lower accuracy than for the other stars.}. As shown in the last column of Table \ref{tab:finalresults}, the rms scatter in the residual is 0.010 mag among the eleven stars (and the average residual is zero). This value reflects the combination of the errors in the $K_{\text{S}}$ value as well as that in the {\it Spitzer} photometry, so it suggests that the uncertainty in the latter ([3.6] measurements relative to Sirius) for individual stars is less than 1\%. Given the excellent color agreement in our first test, a similar uncertainty is expected in the [4.5] measurements.

Yet another comparison uses the DIRBE 2.2 and 3.5 $\mu$m photometry from \citet{smith2004} for Sirius, $\beta$ Gem, and Procyon (the other stars we measured with {\it Spitzer} are too faint for accurate DIRBE photometry). We have made this comparison in two ways: (a) comparing the stars with DIRBE at 2.2 $\mu$m and correcting as in equation (1) to IRAC [3.6]; and (b) comparing the stars with DIRBE at 3.5 $\mu$m with no correction to IRAC [3.6]. For method (a) we have transformed the DIRBE measurement to 2MASS $K_\text{S}$. This was achieved by comparing transformed $K_\text{S}$ and DIRBE photometry for more than 400 stars (see Paper I for the transformations and tests of their accuracy). The difference between the two $K$-band measurements is constant with little color dependence over the relevant range for Procyon and $\beta$ Gem, so all that was required was to adjust the zero point assumed for the DIRBE photometry. The predicted differences relative to Sirius are 0.2095 and 0.211 mag respectively for $\beta$ Gem and 0.664 and 0.670 mag respectively for Procyon, to be compared with the measured values of 0.202 and 0.671. If we average the two determinations for each star from DIRBE, the residual of observed minus prediction is $-$0.008 mag for $\beta$ Gem and +0.004 mag for Procyon. 

All of these tests using other sources of photometry indicate that our {\it Spitzer} PSF wing photometry is accurate to 1\% or better.

\begin{table*}
\tablewidth{0pc}
\begin{center}
\caption{Rayleigh-Jeans Model Predictions \label{tab:rjmodel}}
\begin{tabular}{llccccccc}
\hline
\hline
Name & T$_{\text{eff}}$ &  $\Delta$T$_{\text{eff}}$\tablenotemark{a} & ref.\tablenotemark{b} & $\theta_{\text{DL}}$  &  $\Delta \theta_{\text{DL}}$   &  ref.\tablenotemark{c}  & pred./meas.\tablenotemark{d} & uncertainty \\
  &  (K)  &  (K)  &  &  (mas)  &  (mas)  &  &  [3.6] &  \\
\hline
Sirius	&9800	&	50	&	1	&	6.041	&	0.017	&	1	&	0.983	&	0.007	  \\
Procyon	    &6542 	&	25	&	2	&	5.470	&	.04	&	2	&	0.998	&	0.011	  \\
HD  30652	& 6428	&	25	&	2	& 1.5269	&	0.004	&	3	&	0.978	&	0.006	 \\
HD 102870	& 6161 &	35	&	2	&	1.431	&	0.006	&	3	&	1.002	&	0.008	  \\
HD 142860	& 6278	&	25	&	2	&	1.217	&	0.005	&	3	&	1.009	&	0.007	  \\
HD  19373	& 5958	&	35	&	2	&	1.246	&	0.007	&	3	&	1.013	&	0.010  \\
HD 126660	& 6222	&	25	&	2	&	1.109	&	0.007	& 3	&	0.992	&	0.010	\\
HD 215648	& 6143	&	25	&	2	&	1.091	&	0.008	&  3	& 1.009	&	0.011	  \\
HD 173667	& 6369 &	25	&	2	&	1.000	&	0.006	&	3	&	1.017	&	0.010	 \\
\hline
\end{tabular}
\end{center}
\tablenotetext{a}{Temperature uncertainty are assumed to be 25 K except when the temperature is based on a single spectrum, then 35 K. We adopt a uncertainty of 50 K for Sirius as being a much hotter star compared to others. }
\tablenotetext{b}{References for stellar temperatures are (1) \citet{gebran2016}; (2) \citet{bermejo2013}.}
\tablenotetext{c}{References for the limb-darkened diameters are (1) \citet{davis2011}; (2) \citet{cruzalebes2019}; (3) \citet{boyajian2012}.}
\tablenotetext{d}{The numbers in this column are the magnitude ratios between the predicted 3.6 $\mu$m flux using the Rayleigh-Jeans assumption and our measured one from Table \ref{tab:finalresults}, normalized to 1 on average for the entire sample.}
\end{table*}

\subsection{Stellar SEDs}

Accurate photometry on the Rayleigh-Jeans portion of stellar spectral energy distributions (SEDs) is relatively insensitive to details of the physics of the stellar emission. This behavior is the foundation of the Infrared Flux Method (IRFM) for estimating stellar diameters and deriving an absolute calibration from stellar sources \citep{blackwell1979}. Alternatively, for stars with very accurate diameter measurements through interferometry, infrared measurements become a sensitive test of stellar models. 

It is beyond the scope of this paper to compare our results with detailed stellar models. However we can make a simple test by seeing if the fluxes scale in proportion to the diameter of a star squared (e.g., the solid angle ($\theta_{\text{LD}}$) it subtends) times its temperature (T$_{\text{eff}}$) under the Rayleigh-Jeans assumption, as the very simplest relation physics would predict. Results for this very simplistic model are shown in Table~\ref{tab:rjmodel} excluding $\beta$ Gem (a K giant), HD 10476 (its later spectral type undermines this simple assumption) and HD 165459 (too faint to have direct diameter measurements). It works remarkably well, with flux ratios at IRAC [3.6] for all of the stars relative to Sirius within $\sim$2\% of the predictions of the model, and most of them consistent with the model within the errors from propagation of the uncertainties in stellar temperatures and the diameters. \citet{bohlin2017} have found that modern models of the spectrum of Sirius are consistent with the diameter determined by \citet{davis2011} (and used in our comparison) within the errors, i.e., to within $\sim$1\%. Combined with the results in Table~\ref{tab:rjmodel}, the simplistic Rayleigh-Jeans model appears to be surprisingly accurate.

\section{Conclusion}
\label{sec:conclusion}

We have demonstrated accurate photometry over a dynamic range of eight magnitudes, from Sirius to the A-star HD 165459. To do so, we have exploited the very high optical stability of the Spitzer Space telescope, enabling us to carry out the photometry on the wings of very saturated images of the stars. Our photometry is integrated into a general discussion of the transfer of Sirius photometry to stars sufficiently faint to be measured with modern instrumentation in Paper I as discussed in Section \ref{sec:tests}. Our results agree excellently with those from other methods, with the important feature that our measurement is direct and does not depend on a chain of photometric transfers. As discussed in Paper I, Sirius is an ideal A-star to define photometry zero-points from optical to infrared and preserve a huge body of photometry based on A0V stars. Using the relative values summarized in Table \ref{tab:finalresults}, the absolute fluxes of these stars presented in this study can be easily obtained once the proposed values for Sirius ($V=K_{\text{K}}=$ [3.6] = [4.5] = $-$1.395 mag as presented in Paper I) are fully tested and validated, a goal for upcoming papers.

\facilities{{\it Spitzer} (IRAC)}

\acknowledgments

The authors thank Karl Misselt for helpful conversations and advice during the data analysis. This work is based on observations made with the Spitzer Space Telescope, which is operated by the Jet Propulsion Laboratory, California Institute of Technology. The work was supported by NASA grants NNX13AD82G and 1255094.

\bibliography{ksuref}

\end{document}